# GERBERTVS

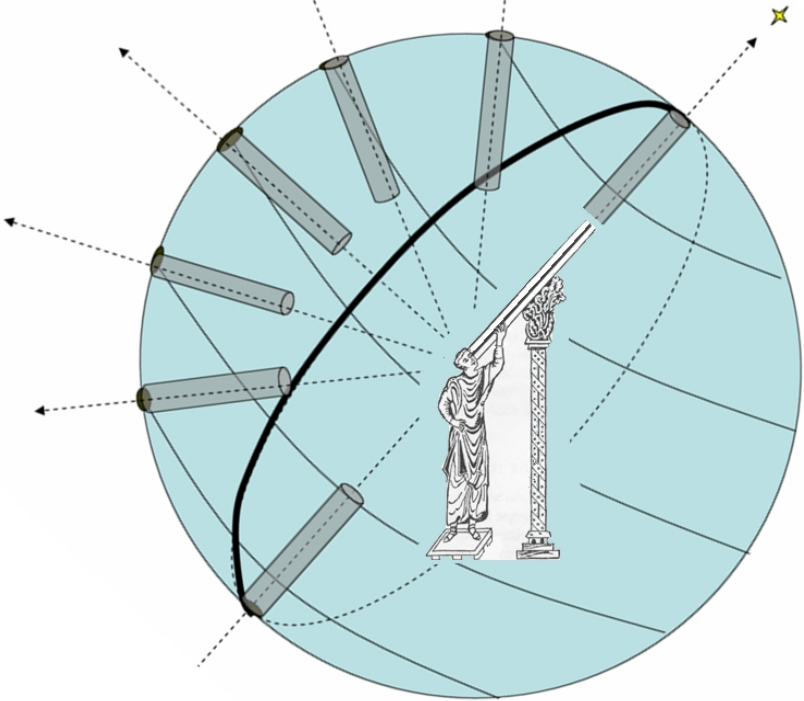

http://www.icra.it/gerbertus

**International academic online publication
on History of Medieval Science
vol. 11/2018**

*Costantino Sigismondi and Hamed Altafi*

# Visual and Hα measurements of solar diameter of 9 may 2016 mercury transit

Costantino Sigismondi *(ICRA/Sapienza &G. Ferraris InstituteRoma)*
Hamed Altafi *(Iranian National Observatory, Institute for Research in Fundamental Sciences, Tehran)*

**Abstract** Visual observations of 2016 Mercury transit ingress made in visible light (Rome) and in H-alpha line (Tehran) are compared to evaluate the quote 1.84" of active H-alpha regions at the solar latitude of the transit, without a confirmation of the theory of an oblate Sun at minimum activity. A variation of -0.12" in the photospheric radius, within 0.17" is found.

**Sommario** Le osservazioni visuali del transito di Mercurio del 2016 sia in luce visibile a Roma che nella riga H-alfa a Tehran sono comparate per valutare la quota di 1.84" delle regioni attive in H-alfa alla latitudine solare del transito, senza la prova che il Sole sia prolato al minimo. Una contrazione di 0.12", entro 0.17" di errore del raggio fotosferico è trovata.

**Instruments and location of the observations**

Visible light: refractor 3″ f/7 used in projection at Rome Sapienza University, with fixed mount: 41° 54′ 12.41″ North, 12° 30′ 48.87″ East, 80 m elevation. 2) H-alpha: Lunt solar telescope with 0.7A band pass on equatorial mount and tracking motion, in Teheran: 35° 44′ 57.62″ North 51° 26′ 39.9″ East, 1399 m elevation and seeing 2″.

The conditions of the sky were clear (thin and rapid clouds 1°/6s passing on the Sun in Rome during the first minute) at both locations at the ingress, for first and second contacts. Both observers already experienced visually Mercury (2003) and Venus (2004 and 2012) transits.

To locate the ingress position before t1 on the solar limb (position angle P.A.=83.3° in Rome) the preceding (East) limb (P.A.=90°) in the projected image with fixed mount was considered, and the sunspot map[1] was compared to the field of H-alpha telescope.

---

1 https://www.solarmonitor.org/?date=20160509

The video of the observation in Rome is published,[2] the appearance of the second contact at the H-alpha telescope is drawed here; the black drop -if any- was instantaneous.

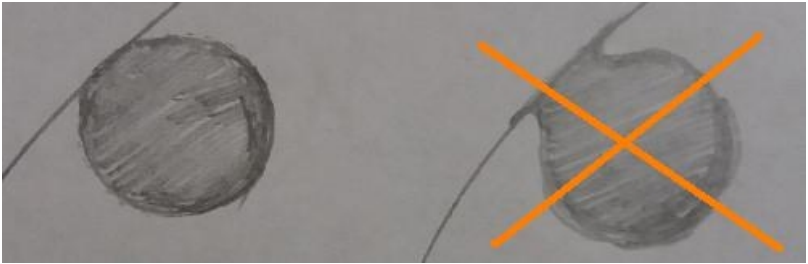

The H-alpha observation was made at 35X magnification and the Ethalon was tuned through pressure adjustment in such way that can see both surface details and limb as sharp as possible, not having emphasis on resolving prominences.

Just after the transit start, the magnification was increased a bit using the zoom eyepiece to see the event more accurately in hope to record probable black drop effect. Mercury was in the center of field of view, as even in very high quality telescopes, off-axis appear some aberrations and optical defects because of the laws of optics.

**Observational data and discussions:**

The following table summarizes the observations

| Location | Observed times [UTC] | Calculated time [UTC] for white light |
|---|---|---|
| Rome | t1 11:13:57 | t1 11:12:10.3(.2) at 65.6° altitude P.A.=83.3° |
|  | t2 11:15:23 | t2 11:15:21.3(.1) |
| Tehran | t1 11:11:17 | t1 11:11:14.4(.3) at 50.1° altitude P.A.=83.1° |
|  | t2 11:13:58 | t2 11:14:25.4(.3) |

The ephemerides used are calsky.org and between parentheses

---

2   https://www.youtube.com/watch?v=4ORAlX7iPKc

() the ones of Helios program. The difference between them is of 0.1 s, corresponding to 6 km in the orbit of Mercury. P.A.=83.3° corresponded to the solar latitude -15.58° for that day (program Helios v3.2).

The observation made in Rome made by projection with several students evaluating their instant of perception for t1 and t2. For t2 a spread of 2 seconds has been found and is recorded in the video, with a case of 26 s of delay, due clearly to the personal eye resolution.

**Theoretical optical resolution and seeing in Rome:** the seeing and the resolution of the instrument+eye can be estimated *a posteriori* by the time of appearance of Mercury's disk: that has been 1m47s later than ephemerides, 107s at 6.3''/100s correspond to 6.75 arcseconds. Mercury had to enter the limb for such dimension (half of its diameter) in order to be seen by us.

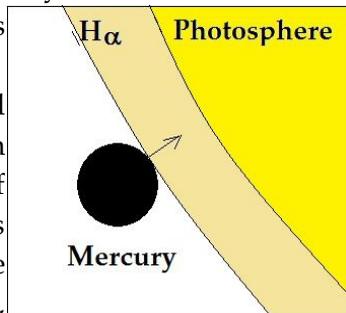

The projected image of the Sun had a diameter of 13 cm and it was seen from 50 cm of distance. The disk of Mercury was 12'' and the Sun was 1901'', so the dimension of the planetary disk was a 0.82mm dot, 338 arcseconds wide at 50 cm of distance, or 5.64 arcminutes. The resolution of the naked eye is averagely 1 arcminute, able to see a dark feature of 0.14 mm on that projected solar disk, a feature corresponding to ϱ=2.13″ in the sky. The theoretical resolution of the telescope is θ=1.22λ/D=1.4 arcseconds according to Rayleigh formula. The atmospheric seeing has been estimated as σ=2.5'' because it permitted to observe clearly the distance between the double spot of the active region NOAA12542[3] which was 5.2''. Combining in quadrature

---

3   https://www.solarmonitor.org/region_pop.php?

the previous factors we obtain a final optical resolution for that event of r=√ (ϱ²+θ²+σ²)=3.6''.

**Improving 20 times the spatial resolution by transit's timing**
The circumstance of the Mercury transit and its timing allowed us to improve of a factor of 20 such theoretical resolution, and applying it to the solar diameter.

We verified that the measurement of t1 in Rome was affected by the "effect surprise" due mainly to the poor contrast of the image in the projected Sun, with respect to a telescopic image.

The second contact t2, conversely, was spotted with only 2s of delay (2-4 s for most of the observers) with respect to the ephemerides: the black drop -if any- was confined within 2s (as occurred for H-alpha observations), and it was an effect of the combination of the previous optical resolutions with the seeing, without the "effect surprise". These 2s of delay of t2 can be attributed to the variation (contraction) of the radius of the Sun of ΔR=-0.12''±0.17'' with the uncertainty given by the quadrature sum of two terms: 0.12'' due to the possible black drop and another 0.12'' due to timing uncertainty.

**Black drop effect:** In addition to 2s of delay of t2 perceived in Rome, the H-alpha observation shown that just after the planet fully entered the Sun's disk, a deformation near the limb was seen (see the drawing in fig. 1); as a possible signature of black drop effect. Any obvious conjunction between the planet's edges and the inner limb of the Sun was not seen.

The observer was expecting this effect lasting more seconds, but only no more than 3s after what has been called black drop, the planet edge was clearly separated from the solar limb and this was defined as the second contact time.

**Height of H-alpha region with respect to the photosphere**
The first contact t1 in H$\alpha$ was observed 2.6s later than

date=20160509&type=chmi_06173®ion=12542

photospheric radius' ephemerides. The second one t2 27.4s before than predictions for photospheric radius. Mercury was perceived entering the H$\alpha$ layer of the Sun with the opposite limb 2 minutes and 41s=161s after the t1. Since H$\alpha$ activity is at higher quote with respect to the photosphere, a general advance of t1 and t2 should be expected, but no difference in $\Delta t=t2-t1$, because the curvature of the H$\alpha$ layer is similar to the photosphere (fig. 2), so the interval $\Delta t=t2-t1$ from ephemerides is 3 minutes 11s, 191.0s in both wavebands. Moreover there is no significant absolute error >0.1s on modern ephemerides.

The program Helios developed in Brazil by Helio de Carvalho Vital gives for the same location just 0.1s offset with respect to Calsky (Switzerland), corresponding averagely to 6 Km in the orbit of Mercury at its average speed. The atmospheric seeing $\sigma$=2'' contributed to the "suprise effect" for t1 timing by 30s.

The diameter of Mercury was 12.05'' and its entrance velocity was 6.309''/100s, say 2'' in 31.7s: the agitation of the observed solar limb profile was larger than the portion of Mercury already entered.

Similarly to Rome observations, the timing t2 of the second contact, was unaffected by the black drop and by "surprise".

The ingress phase in H$\alpha$ then reliably started at t1 corrected by 31.7s: t1corr=15:40:45.3, 29.1 s before the predicted photospheric contact time. The H$\alpha$ radius was therefore 1.84'' larger than the 959.63'' photospheric standard radius. Consistently with previous considerations the final H$\alpha$ radius at solar latitude -15.58° is 1.84''±0.19'' above than the photospheric one, being the 0.19'' the 3s of maximum allowed black drop.

**Prolate H alpha solar atmosphere and planetary transits**

The are indications in literature (Filippov and Koutchmy, 2000; Filippov, Koutchmy and Vilinga, 2007) on the prolateness of solar chromosphere at solar minimum, disappearing at maximum. In 1996 the height of the chromosphere at the center of Hα line was ranging from 4.3-4.4 Mm=6'' at the equator to 6Mm=8.3'' at the poles (Johannesson and Zirin, 1996). In the following table we compare such data with the ones obtained during planetary transits by Sigismondi (2015).

| Date | 11.4 year cycle phase 0°=1/1/2008 | Solar Latitude | Quote H alpha layer |
|---|---|---|---|
| 1996.6 | 0° | 0° | +6'' (J&Z96) |
| 1996.6 | 0° | 90° | +8.3'' (J&Z96) |
| 2003, 7 May<br>2006, 8 Nov | 218°<br>323° | 51.3° North<br>28.1° South | +0.41'' (678 nm) |
| 2004, 8 June | 250° | 44.1°±0.8° South | +0.39''±0.01'' |
| 2012, 6 June | 142° | 36.5°±0.8° North | +0.27±0.01'' (617.3 nm)<br>+0.28±0.16'' (551±88 nm) |
| 2016, 9 May | 268° | 15.6° South | +1.84''±0.19'' |
| 2004, 8 June | 250° | 44.1°±0.8° South | +0.49''±0.01' (678 nm)' |

How to explane these data in a single framework?

From this table no trend is evident; and the question of the tuning of H-alpha filters appears relevant on the H-alpha limb definition, since the data of 1996 are a factor of 4 larger than others. Moreover an aperture of 60 mm (2004) and the one of 100 mm (2016) for the observation of the limb in H-alpha give respectively 1.7'' and 1.0'' optical resolution; and 2.8 times in intensity with a possible modified cutoff with respect to the background luminosity.

The Hα observations for 2004 Venus transit[4] were made by Anthony Ayomamitis with a Coronado 60 mm and 0.7Å bandpass filter combined with a Televue refractor. The Hα quote was obtained from a sequence of such timed photos as 387±9 milliarcsec and from the interpolation of satellite data: 414±11 milliarcsec (Sigismondi, et al. 2015).

The interpolation of 411 mas for extra Hα radius with respect to photospheric one, has been made between data from 2003-2006 transit of Mercury at 676.78 nm with SOHO satellite, and the ones at 617.3 nm SDO data of the transit of Venus 2012, but they give only the continuum while Hα cannot be considered as the continuum's extrapolation at 656 nm, it is another physical process; moreover the Sun was in different stages of its 11-year cycle.

**Variations of Hα quote by tuning the filter**

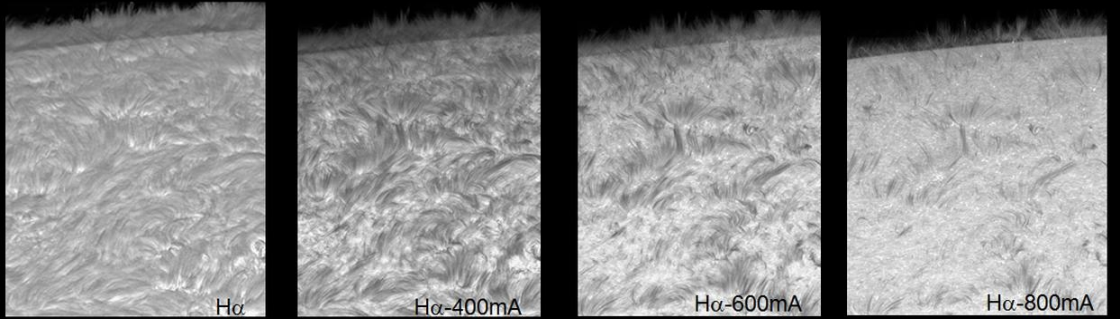

From this sequence of Dutch Open Telescope images around the center of Hα line, we see the heigth of the forest of spicules decreaseing to half value toward 0.8Å off-center. The spicules are features of 0.7'', and cannot explain variations of 8'' (or 2.3'' from pole to equator) found in the above table, unless the prolate Hα Sun is strongly dependant on the particular cycle.

**Conclusions on the solar diameter variations**

The chromosphere thickness is about 2500 km, corresponding to 3.45'' at 1AU; the spicule arrive at 500 km (0.69'') and are visible in H$\alpha$ (Bialkow coronograph); the H$\alpha$ layer is within these dimensions, and its local extention can be measured visually during planetary transits at given phase of solar cycle. The hypotesis of a prolate Sun during solar minimum (0° phase) with a difference of 2.3" between Polar and Equatorial radius is out of our data range, being 268° or -92° the phase in 2016 Mercury transit, of a cycle, the 24, fainter than the 23.[5]

---

4   http://www.perseus.gr/Astro-Planet-Ven-Tr2004.htm

Interpolating only H$\alpha$ data from table 1 at solar latitude 15.6° South, we have: phase 0° (1996.6) +6.4'', phase 250° (2004.5) +0.4'' at 44°, phase 268° (2016.4) +1.8''. Being more precise the last two measurements made during transits, we can see an increase of +1.4'' in the H$\alpha$ radius from cycle 23 to 24.

H$\alpha$ radius is 1.84''±0.19'' above the photospheric one, which is 0.12''±0.17''smaller than the standard one.

The scientific value of historical accurate visual observations like Gambart's one in Marseille on May 5, 1832 is confirmed by this analysis of present visual data, and it is relevant for studying the long term evolution of the solar diameter.

**Acknowledgments**

To Serge Koutchmy and Pawel Rudawy for the explanations on the H-alpha Sun.

---

5  https://wattsupwiththat.com/2018/03/18/approaching-grand-solar-minimum-could-cause-global-cooling/